\documentclass[]{article}
\usepackage[OT4]{fontenc}
\usepackage{hyperref} 
\usepackage{graphicx} 
\usepackage{psfrag}
\usepackage{subfig}
\usepackage{authblk}

\begin{document}

\title{Heider balance, asymmetric ties, and gender segregation}

\author[a]{Ma{\l}gorzata J. Krawczyk}  
%% \email{gos@fatcat.agh.edu.pl}
\author[b]{Marcelo del Castillo-Mussot}  
%% \email{gos@fatcat.agh.edu.pl}
\author[b]{Eric Hern\'andez-Ram{\'\i}rez}  
%% \email{gos@fatcat.agh.edu.pl}
\author[b]{Gerardo G. Naumis}  
\author[a]{Krzysztof Ku{\l}akowski \thanks{kulakowski@fis.agh.edu.pl}}
% \ead{kulakowski@fis.agh.edu.pl} 

\affil[a]{AGH University of Science and Technology,
Faculty of Physics and Applied Computer Science,
al.~Mickiewicza~30, 30-059 Krak\'ow, Poland.
}

\affil[b]{Instituto de F{\'\i}sica, Universidad Nacional Aut\'onoma de M\'exico, 04510 M\'exico D.F., M\'exico
}
\providecommand{\keywords}[1]{\textit{Keywords:} #1}
\providecommand{\pacs}[1]{\textit{PACS:} #1}
\maketitle

\date{}

\begin{abstract}
To remove a cognitive dissonance in interpersonal relations, people tend to divide our acquaintances into friendly and hostile parts, both groups internally friendly and mutually hostile. This process is modeled as an evolution towards the Heider balance. A set of differential equations have been proposed and validated (Kulakowski {\it et al}, IJMPC 16 (2005) 707) to model the Heider dynamics of this social and psychological process. Here we generalize the model by including the initial asymmetry of the interprersonal relations and the direct reciprocity effect which removes this asymmetry. Our model is applied to the data on enmity and friendship in 37 school classes and 4 groups of teachers in M\'exico. For each class, a stable balanced partition is obtained into two groups. The gender structure of the groups reveals stronger gender segregation in younger classes, i.e. of age below 12 years, a fact consistent with other general empirical results.\\
\\
\end{abstract}

%\keyword s{Athermal phase transition; Crowd dynamics; Long-range interactions; Computer simulation}

%\begin{keyword}
%networks \sep dynamics \sep communities \sep gender
%\PACS 05.45.-a \sep 89.65.Ef \sep 89.75.Fb 

%\end{keyword}
%\end{frontmatter}

%% ===========================================================================
\section{Introduction}
%% ===========================================================================
The gender segregation is a well-known phenomenon which shapes our social life from the preschool years up to puberty \cite{macc}. It becomes visible already between children of age less than three years. Estimations of frequency of overall pro-social attitude of children of 33 months old indicate that it increases systematically when related to the same-sex children: for boys (from 19 vs girls to 29 vs boys) and girls (from 17 vs boys to 31 vs girls) \cite{jack}. When children grow, the effect is reinforced; the proportion of time spent with the same sex to cross-sex partners increases from 3:1 for the children of 4.5 years to 11:1 for the children of 6.5 years \cite{maja}. While the segregation is well documented, its origin remains disputable; for an extensive discussion we refer to \cite{macc,jack,maja,hoff,meht} and references cited therein. In particular, cognitive factors have been distinguished as one of possible driving forces towards the segregation \cite{macc,jack}; this approach assumes that children (even very young) are able to 
distinguish 
sex of their coplayers and their own. Another possible explanation is that boys and girls have different styles of play, respectively more aggressive and more passive, and they prefer to contact with those of the same style. As none of these theories has been confirmed in experimental studies \cite{jack,hoff,moll}, the issue remains open; yet the segregation persists under different forms in adulthood \cite{meht}. What is undisputed is that it is a group effect, i.e. children use sex (or gender) as a criterion of social grouping \cite{macc}. We note that after \cite{macc}, we use the words {\it sex} and {\it gender} interchangeably.\\

Here we are interested in a general method of detection of the gender segregation by means of data analysis. We assume that the input data are given in the form of a matrix, which contains information on preferences of agents of a given group towards other members of this group. Data in this form has been collected and analysed many times, with the paper of Zachary on the karate club \cite{zach} as a canonical example. The new element here is that we allow for asymmetric relations, i.e. the feeling of person A about person B is not necessarily the same as the feeling of B about A. In other words, the related matrix is asymmetric, and the related networks of relations are directed.\\

Focusing on social relations, when designing an appropriate method of data analysis, our starting point is the model of the change of perception: the removal of cognitive dissonance \cite{ijmpc,grax,aap,cse}. As the outcome, the method provides the group partition into two mutually hostile parts, with friendly relations within each part. This state is equivalent to a consistent partition of each triad of agents into a pair of friends and their common enemy; alternatively, all three ties in the triad are friendly. The state is known as the Heider balance \cite{hei1,hei2}. Per analogy to an Ising spin glass \cite{fihe}, in the balanced state there is no frustrated triads; the product of bonds is positive for each triad.  \\

Our model approach \cite{ijmpc,aap,cse} applies a set of differential equations and it has been successfully validated with using selected sociological data \cite{zach,fre}, yet all of them contained only symmetric relations. As was demonstrated in \cite{stro}, our method is free from stable unbalanced (so-called 'jammed') states, which plague discrete methods based on the Monte-Carlo scheme \cite{ant1,ant2}. The goal was not only to obtain the final partition, but also to simulate the actual dynamics of the cognitive process. It is important to note, that the interpretation of the obtained partition depends on the context. In particular, the data analysed in \cite{fre} contain just 
the list of attendance of a set of persons on their meetings, and the quality of relations is represented by the correlations of this attendance. In this case, the persons involved could be ignorant about any hostility or conflict in the group. In general, the method allows only to indicate two groups, with internal relations more tight (or friendly), than between the groups.\\

Our goal here is to generalize our former method to the case of non-symmetric relations. The motivation is twofold. The first is obvious: basically there is no reason to expect that the relations are symmetric, and actually -- as is shown below -- we are going to cope with data where they are not. The second aim is again to include the psychological process of direct reciprocity, one of the main mechanism of triggering cooperation between otherwise selfish agents \cite{nowak}. With this generalized model, we are going to capture the dynamics of gender separation process, where both contacts in dyads and third-person reactions allow to reproduce a partition into two coherent groups, and where differences in patterns of behavior accepted in these groups stabilize the partition. \\

The generalized equations are applied to the weighted data on mutual liking and aversion, collected in 37 classes in Mexican schools and four groups of teachers. The age of the respondents in the classes varied from 9 to 23. Each individual had to indicate five persons which she or he liked most, and grade them with 1 to 5 points; the same was asked about disliking, with negative grades. These data are presented in the form of non-symmetric matrices, one matrix per class. As a rule, these data do not provide a coherent partition of a class into two sets. This is done with the model differential equations. The outcome can be understood as the partition most close to the initial data. The analysis of the gender distribution of the obtained partition allows to indicate the classes where the gender is meaningful as a criterion of the partition. This is done by the calculation of the correlation between the proportion of males and females in the obtained groups.\\

The next section is devoted to the model equations. Part of this chapter is relegated to the Appendix. There we show that when the equations in their previous form are applied to non-symmetric data, they lead to jammed states, similarly to the non-deterministic algorithms \cite{ant1,ant2}.  Coming back to the Section 2, we demonstrate, that the jammed states are removed in the generalized model, except the case when the rate of the new process is too small. 

In Section III we provide the details on the collected data, which describe the relations between children in 37 school classes. Separate data sets for four groups of adults are taken as a reference point.  In section IV we describe the numerical results of the application of the model equations to the relations between children. The analysis of the results allows to evaluate the level of the gender classification in each class. The last section is devoted to the discussion.

\section{The model}
In the initial formulation \cite{ijmpc}, the time evolution of an element $x(i,j)$ of the symmetric matrix of relations was given by
\begin{equation}
\frac{dx(i,j)}{dt}=G(x(i,j))\sum_{k}^{N-2}x(i,k)x(k,j)
\label{eq1}
\end{equation}

where the role of the term $G(x)=1-(x/R)^2$ was to limit the values of $x$ to a prescribed range $(-R,R)$.

The idea behind Eq. \ref{eq1} is as follows. Recall that the conditions of the removal of cognitive dissonance are as follows: $\it {i})$ friend of my friend is my friend, $\it {ii})$ friend of my enemy is my enemy,  $\it {iii})$ enemy of my friend is my enemy, $\it {iv})$ enemy of my enemy is my friend \cite{hei2}. Let us consider a particular relation $x_{ij}$ between individuals $i$ and $j$. For all their common neighbours $k$, the relations $x_{ik}$ and $x_{kj}$ are either of the same or different sign. In the former case, these two relations are either both friendly or both hostile; in both cases, the triad $(i,j,k)$ is balanced iff $x_{ij}$ is positive. Then, the positive sign of the product $x_{ik}x_{kj}$ is an incentive to improve the relation $x_{ij}$. Conversely, the negative sign of the same product should lead to a deterioration of the relation $x_{ij}$. In total, the time derivative of  $x_{ij}$ is a sum of the products $x_{ik}x_{kj}$  over all neighbors of the pair $(i,j)$.

The application of Eq. \ref{eq1} to the input data, described in the next section, converts each set of relations to a stable partition of the group into two subgroups. Although most of the relations given in the form of the matrix elements are initially zero, in the final partition all of them are equal to $\pm 1$. These non-zero values reflect the important assumption of the Heider balance: in equilibrium, no neutral relations exist \cite{ijmpc,grax,aap}.

We note that the value of $R$ is not relevant for stable states, as $R$ can be incorporated to the time variable when dividing the whole equation by $R$ and using variables $y=x/R, \tau=Rt$. Another option is to put $G(x)=\Theta(x)\Theta(1-x)$. As was shown for instance in \cite{ijmpc}, in a generic case the time evolution drives the matrix elements $x(i,j)$ to their limit values $\pm R$. Then, basically Eq. \ref{eq1} can be written with $G(x)=1$, if the time evolution is numerically halted when the limit values $x(i,j)=\pm 1$ are attained. In \cite{stro}, the factor $G(x)$ did not appear at all, which reduces Eq. \ref{eq1} to a matrix Ricatti equation and allows for an analytical solution. With this condition, the proof that Eq. \ref{eq1} does not produce jammed states, given in \cite{stro}, is valid at least for an initial stage of the evolution.\\

In the case when the input matrix $x(i,j)$ is symmetric, it remains symmetric during the time evolution. Yet, for a non-symmetric input, jammed states appear again. We made an attempt to identify them for small groups, namely $N=3$ and $N=5$, where $N$ is the number of agents in the group. (For $N=4$, the number of terms on the r.h.s. of Eq. \ref{eq1} is even, which makes the analysis more complex, as the terms could cancel to zero.) Namely, we ask if there are stable states which are not balanced. The condition of stability is that for each pair $(i,j)$ such that $i\ne j$

\begin{equation}
x(i,j)=sign(\sum_{k}^{N-2}x(i,k)x(k,j))
\label{eq2}
\end{equation}

The condition of Heider balance is that for each triad of different agents $(i,j,k)$

\begin{equation}
x(i,j)x(i,k)x(k,j) = 1
\label{eq3}
\end{equation}

We should add that for non-symmetric matrices, the order of indices does matter. For example, if the  choice in Eq. \ref{eq3} is $x(i,j)x(i,k)x(k,j)$, the r.h.s of Eq. \ref{eq1} is the same for $x(i,j)$ and $x(j,i)$, which is not our intention. Accordingly, the time derivative of $x(i,j)$ in Eq. \ref{eq1} is $x(i,k)x(k,j)$, and not $x(i,k)x(j,k)$.\\

In the Appendix we show that for $N=5$ nodes, the jammed states do exist. We suppose that this is true also for larger values of $N$. Yet, we also show there that for the network of $N=3$ nodes, there is no jammed states.

In the generalized model proposed here, the equations are

\begin{equation}
\frac{dx(i,j)}{dt}=\alpha (x(j,i)-x(i,j)) +\frac{1-\alpha}{N-2}\sum_{k}^{N-2}x(i,k)x(k,j)
\label{eq4}
\end{equation}

where $\alpha$ is a parameter which measures the rates of two processes; the variation of the relation $x(i,j)$ due to the direct reciprocity (rate $\alpha$) and the variation due  to an influence of a third-person (rate $1-\alpha$). The first term is new; its action is just to gradually eliminate the difference between $x(i,j)$ and $x(j,i)$. It is clear that under its action, the symmetry is restored, if only the value of the parameter $\alpha$ is large enough. Our experience for the Mexican data mentioned in Section III as well as for random matrices is that $\alpha=0.2$ is sufficient to make the state symmetric. Above this value, the first term in Eq. \ref{eq4} gradually vanishes in time and any stable solution is balanced \cite{ijmpc,stro}. We note that although the factor $G(x)$ is absent in Eq. \ref{eq4}, the values of the matrix elements $x(i,j)$ are kept numerically in the range $(-1,1)$. Let us add that for $\alpha =1$, the time evolution reduces to a set of independent pairs $(x(i,j),x(j,i))$, which 
just converge, each pair to identical values $(x(i,j)+x(j,i))/2,(x(i,j)+x(j,i))/2$. The coefficient $N-2$ in the denominator of the second term is just to make an interpretation of $\alpha$ less dependent on the system size. The last remark is that if the first term in Eq. \ref{eq4} is substituted just by $\alpha x(j,i)$, the balance is not attained.\\

\begin{figure}[!hptb]
\begin{center}
\includegraphics[width=.8\columnwidth, angle=0]{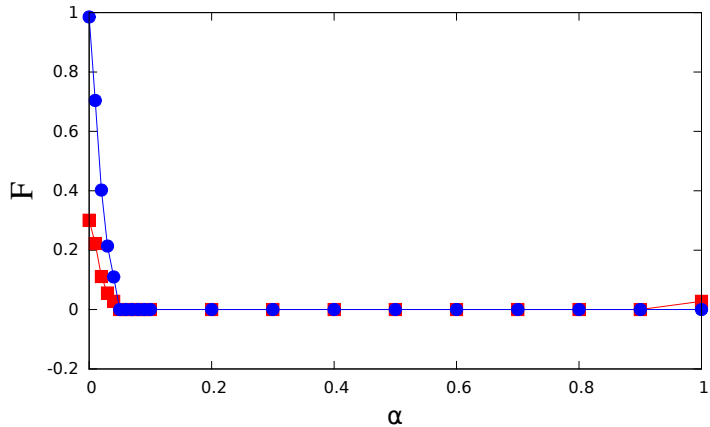}
\caption{A typical dependence of the fraction $F$ of pairs of relations which remain asymmetric ($x_{ij}\neq x_{ji}$) in the stable state (blue circles) and the number of unbalanced triads (red squares) on the parameter $\alpha$. As a rule, the former measure vanishes for $\alpha >$ 0.2; so does the latter, except the range $\alpha >$ 0.9.}
\label{fig:alf}
\end{center}
\end{figure}

In Fig.\ref{fig:alf}, we show a typical dependence of the final state on the parameter $\alpha$; the obtained pattern is reproduced for all investigated classes, as well as for the random non-symmetric input data. For $0.2<\alpha <1$, both the number of unbalanced triads and the asymmetry of the matrix elements do vanish. In all but some cases, the obtained partition of the group does not depend on $\alpha$, if only $\alpha$ remains in the above range. Also, we do not observe any cyclic variations of $x(i,j)$, found in \cite{przem}; we conclude that these oscillations are not generic. 

The conclusion of this section is that the new term in Eq. \ref{eq4}, the one proportional to $\alpha$, allows to remove the asymmetry if only $\alpha>0.2$, i.e. if only the rate of the related process is sufficiently large. There, the fraction of asymmetric links is zero, as shown in Fig. \ref{fig:alf}. Simultaneously, once the equilibrium state is symmetric, the jammed states are removed. This is shown also in Fig. \ref{fig:alf}: the number of imbalanced triads is zero in the equilibrium state, except for $\alpha>0.9$, where the balancing process (rate $1-\alpha$) is too slow.

\section{The data}

We conducted friendship and animosity surveys in 37 classrooms groups  in the Mexico City Metropolitan area with students of different ages and levels. In each group students were asked to rate and order some of their peers in their classroom in two lists. In the first list each student wrote their 5 best friends in order, starting from the best friend, and in the second list they wrote the 5 worst acquaintances or enemies, starting from the worst of them. In this way we constructed 37 networks of interlinked  friendship and animosity within each group. For each person we can sum the total arrows or directed links of either friendship or  animosity to get  useful information.  A more complex description is obtained by assigning values to each arrow depending of the position in the survey lists. For example, values can be assigned to run  from 5 (to the maximum positive feeling) to -5 (the maximum negative feeling). Then a person can get, for instance, a high value as a friend in two ways. First, if he or she 
is chosen as an 'average  friend' by many peers,  or else, he or she is considered to be  a 'very good friend' by just a few peers.  In general, there were some students who were able to mention 5 friends, but mention less than 5 'foes'.
Some general features were found in almost all networks, such as strong differences between friendship and animosity in the way the links are distributed.  In order to compare the distributions obtained in classrooms with a random model network, we performed computer simulations to distribute 5 links randomly of two types in each group.  Then, as in the study of three elementary schools in Yucat\'an, M\'exico \cite{hue}, we found that friendship and animosity statistical distributions exhibit different behavior.

In our case we found that the friendship networks are distributed more randomly, while the animosity networks tend to be more concentrated on fewer persons. That is, the animosity links are focused more on less members, which is an interesting fact. It is important to mention that here we analyze these results from the descriptive and statistical point of view, so further work is needed to understand the causes of the general difference in the asymmetrical distribution of friendship and animosity links in the surveyed groups. 

\begin{figure}[!hptb]
\begin{center}
\includegraphics[width=.5\columnwidth, angle=0]{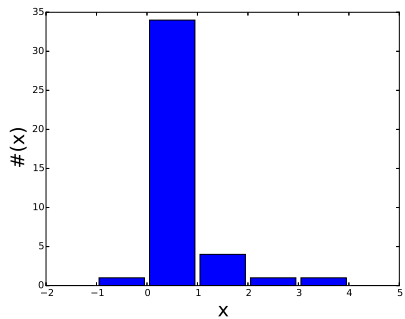}
\caption{The histogram of relations directed towards individuals in the collected data.}
\label{fig:histAll}
\end{center}
\end{figure}

A preliminary analysis reveals that despite the conditions of the poll, the data consist of more positive than negative feelings. This effect is visible in Fig.\ref{fig:histAll}. 

\begin{figure}[!hptb]
\begin{center}
\begin{tabular}{cc}
\includegraphics[width=.5\columnwidth, angle=0]{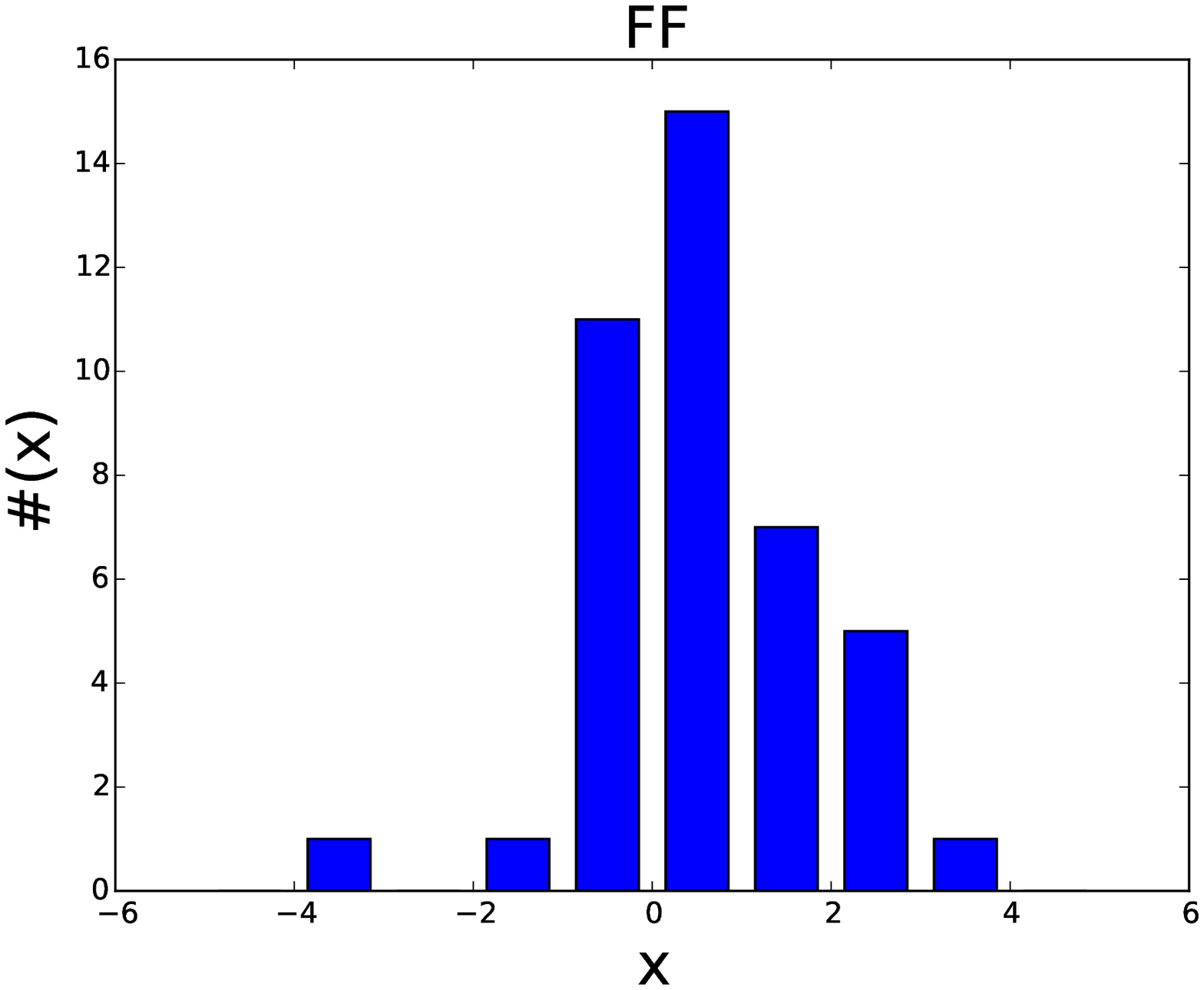}
&
\includegraphics[width=.5\columnwidth, angle=0]{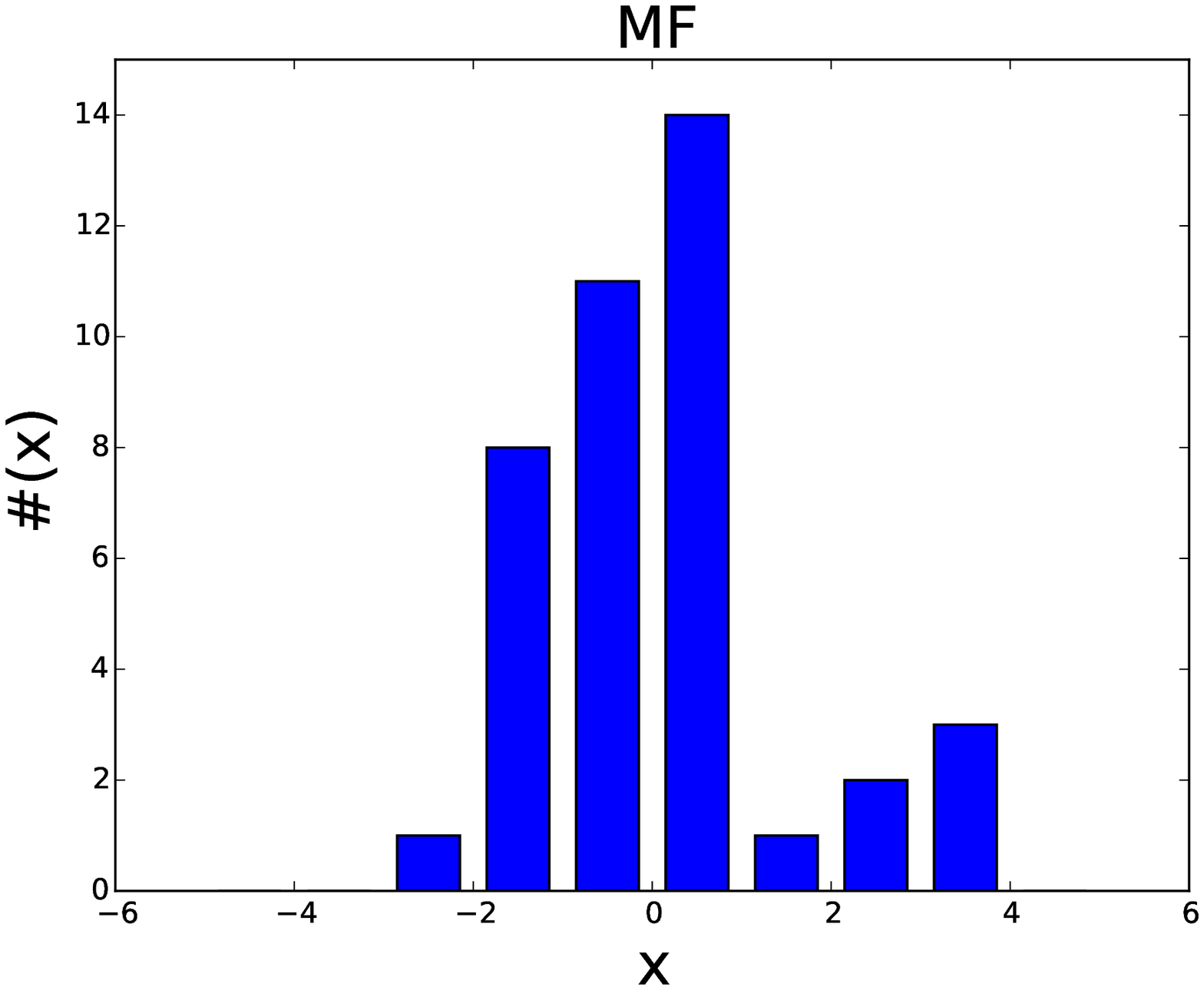}
\\
\includegraphics[width=.5\columnwidth, angle=0]{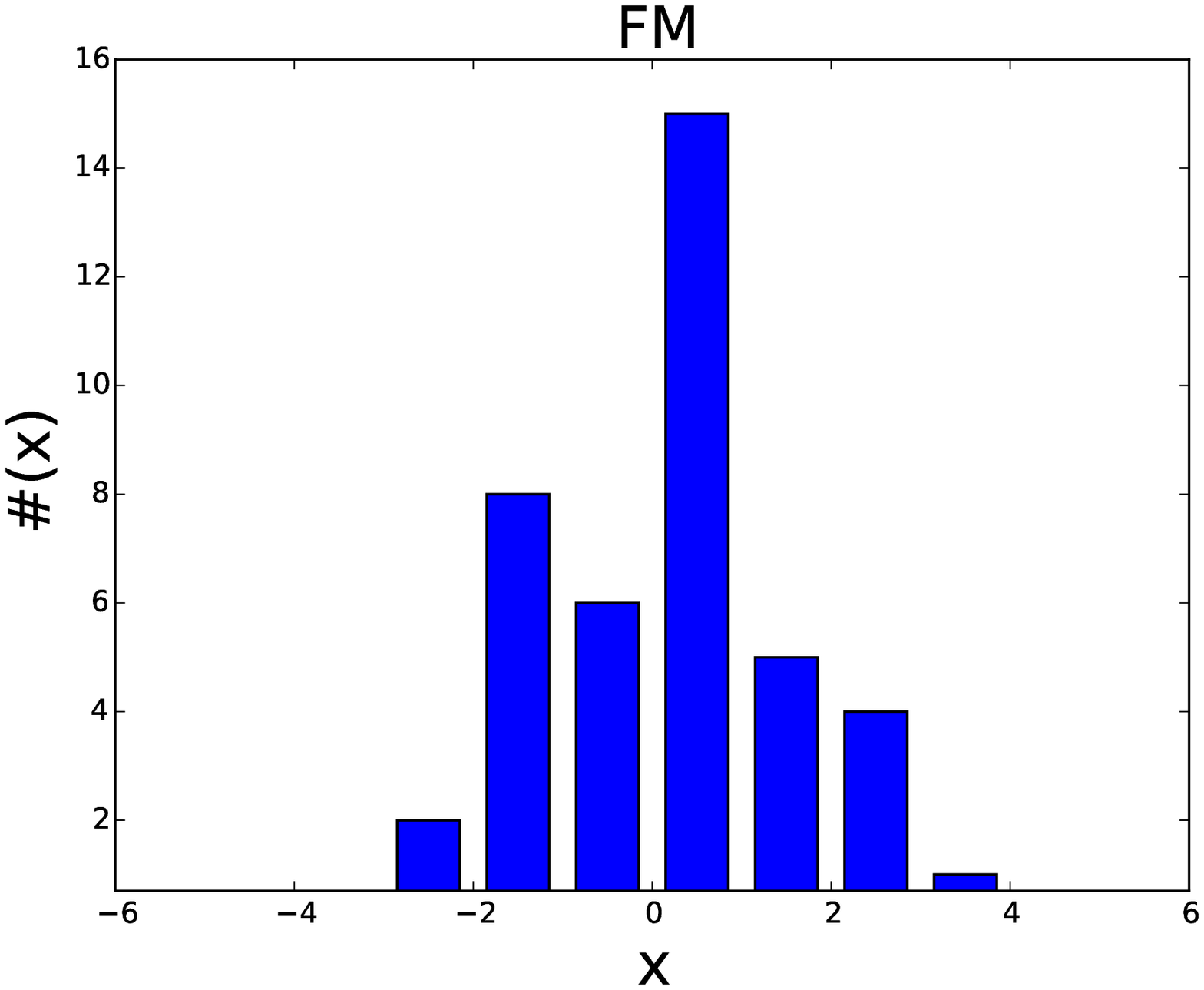}
&
\includegraphics[width=.5\columnwidth, angle=0]{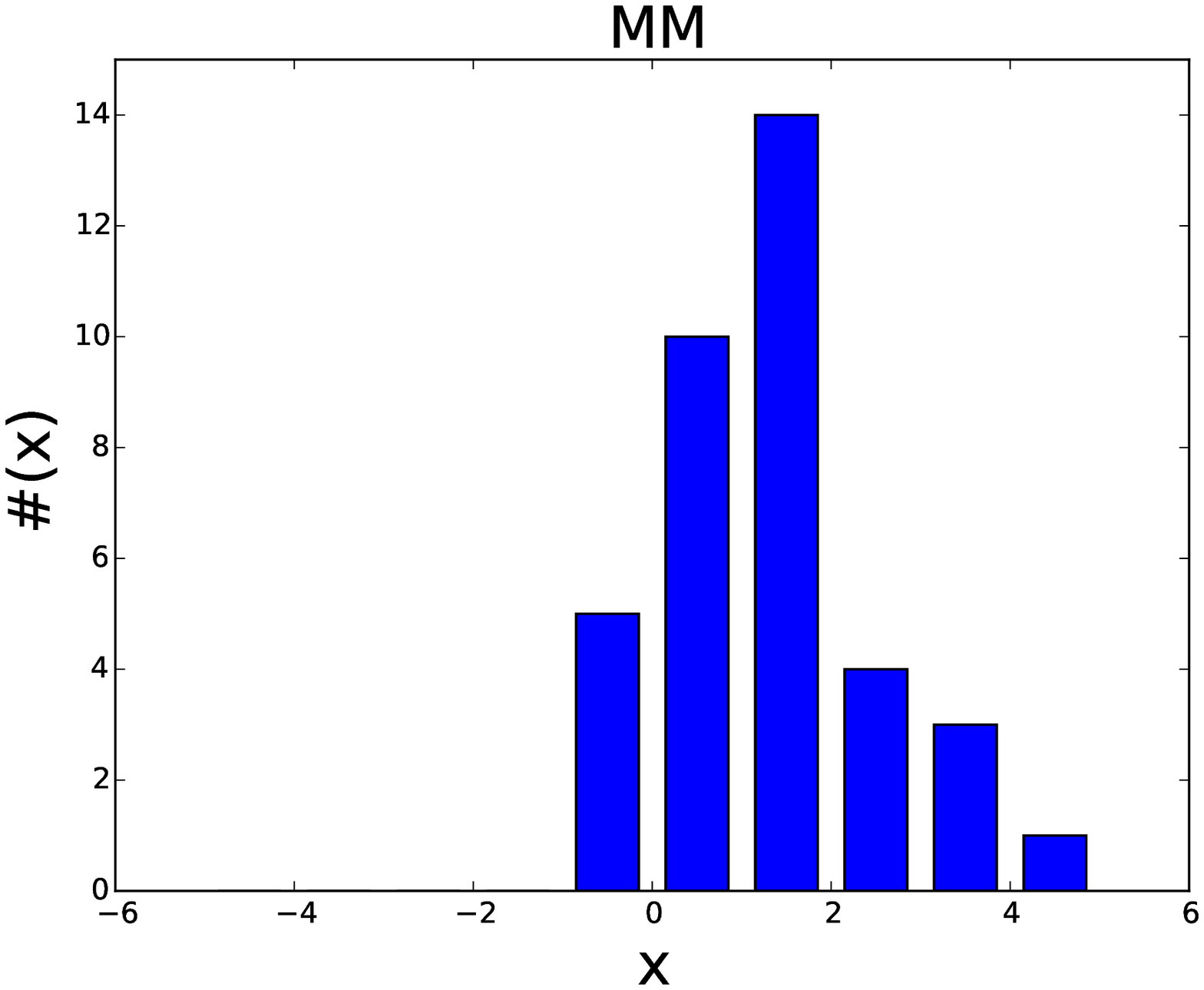}
\end{tabular}
\caption{The histograms of relations directed towards individuals in the collected data, separated with respect to the gender: girls towards girls (top left), boys towards girls (top right), girls toward boys (bottom left) and boys toward boys (bottom right).}
\label{fig:hist}
\end{center}
\end{figure}

Also, there are some differences between preferences of girls and boys. For each person, results are gathered on how he or she is liked or unliked. In Fig.\ref{fig:hist} we show a series of histograms, where these data are gathered. The mean values and standard deviations are calculated for girls about girls, girls about boys etc. The results are close to the overall mean ($<x>\pm \sigma =0.5 \pm 0.7$), with perhaps one exception: the result on boys about boys  $(1.2 \pm 1.1)$, which is slightly more positive than the others. 

In  the next section, the collected data are used as initial conditions for the system of differential equations, which provide the time dependence of particular relations $x_{ij}$.

\section{The results}

\begin{figure}[!hptb]
\begin{center}
\includegraphics[width=.8\columnwidth, angle=0]{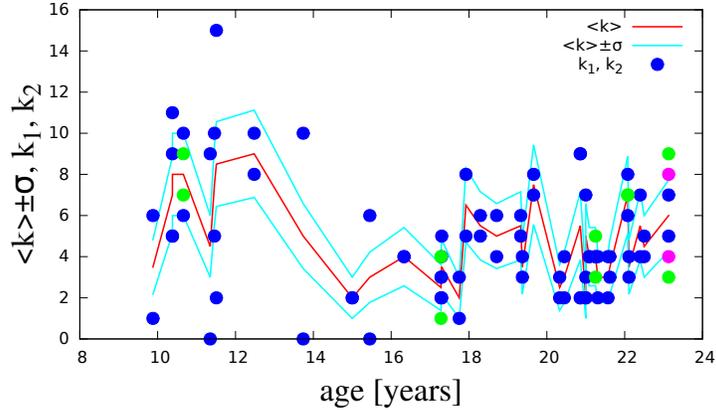}
\caption{For each class, the mean size of the female group is found as the number $k$ of females divided by two (red line). Two green lines above and below the red line mark the mean $<k_1>=k/2$ plus and minus the standard deviation $\sigma =\sqrt k /2$, calculated for the distribution $P(k_1;k)$ given in Eq. \ref{eq5}. The calculated values of $k_1$ are given by blue circles. In these rare cases, when the obtained partition depends on $\alpha$, green circles are added. The mean age of the class is in the horizontal axis.}
\label{fig:fimale}
\end{center}
\end{figure}

\begin{figure}[!hptb]
\begin{center}
\includegraphics[width=.8\columnwidth, angle=0]{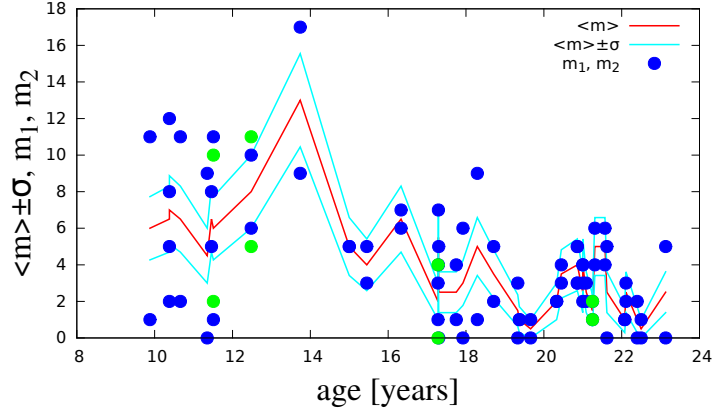}
\caption{For each class, the mean size of the male group is found as the number $m$ of females divided by two (red line). Two green lines above and below the red line mark the mean $<m_1>=m/2$ plus and minus the standard deviation $\sigma =\sqrt m /2$, calculated for the distribution $P(m_1;m)$, the same as given in Eq. \ref{eq5}. The calculated values of $m_1$ are given by blue circles. In these rare cases, when the obtained partition depends on $\alpha$, green circles are added. The mean age of the class is in the horizontal axis.}
\label{fig:male}
\end{center}
\end{figure}

For each class, including four groups of teachers, and for each gender separately, we calculated the estimated content of males (boys, men) and females (girls, women) in two parts. Obviously, the estimated content of each gender in each part is $k/2$ ($m/2)$, where $k$ ($m$) is the actual number of females (males)  in a given class.  The related standard deviations $\sigma$ are calculated from the distribution

\begin{equation}
P(k_1;k)=2^{-k}\frac{k!}{k_1!(k-k_1)!}
\label{eq5}
\end{equation}
where $k_1$ is the number of females in the class. Accordingly, the standard deviations are $\sigma =\sqrt{k}/2$; the same formula for $P(m_1;m)$ is used for males.  The actual, i.e. calculated values of $k_1$ and $m_1$ are obtained from the numerical solution of Eq. \ref{eq4}, with the data collected in polls as the initial values of the matrix elements. In Figs.\ref{fig:fimale} and \ref{fig:male} we show these data together. In Fig.\ref{fig:fimale}, three lines mark the obtained values of $<k_1>-\sigma$, $<k_1>$ and $<k_1>+\sigma$, from below to the top, against the mean age of females in the class. The same for males is done in Fig.\ref{fig:male}. 

These data allow to estimate, if the deviations of $k_1$ and $m_1$ from their mean values $<k_1>$ and $<m_1>$ are statistically meaningful. Namely, if the obtained numbers of girls $k_1,k_2$ (Fig. \ref{fig:fimale}, blue circles) and boys $m_1,m_2$ (Fig. \ref{fig:male}, blue circles) in the groups are more far from the estimated mean values (Figs. \ref{fig:fimale} and \ref{fig:male}, red lines) than the standard deviations (Figs. \ref{fig:fimale} and \ref{fig:male}, green lines), these numbers can likely be assigned to the gender segregation. The results shown in Figs. \ref{fig:fimale} and \ref{fig:male} indicate, that the segregation appears for young children. Yet, these data do not capture the correlations within the parts of the class. This correlation can be semi-quantitatively evaluated as 

\begin{equation}
J(p)=\frac{(k_1(p)-<k_1>(p))(<m_1(p)>-m_1(p))}{\sigma_k(p) \sigma_m(p)} 
\label{eq7}
\end{equation}

where the index $p$ runs through the 37 classes and the 4 groups of teachers. The quantity $J(p)$ is a measure of how the surplus of males in one 'camp' of a class $p$ is correlated with a surplus of females in the other 'camp' in the same class $p$. If the number $m$ of males and the number $k$ of females happens to be the same in each class, then $J(p)$ is just the contribution to the Pearson correlation between the number of females and the number of males, with minus sign. However, in our data both $m$ and $k$ vary from one class to another. For the distribution given in Eq. (\ref{eq5}), we get

\begin{equation}
J(p)=\frac{(2k_1(p)-k(p))(m(p)-2m_1(p))}{\sqrt{k(p)m(p)}}
\label{eq9}
\end{equation}

\begin{figure}[!hptb]
\begin{center}
\includegraphics[width=.8\columnwidth, angle=0]{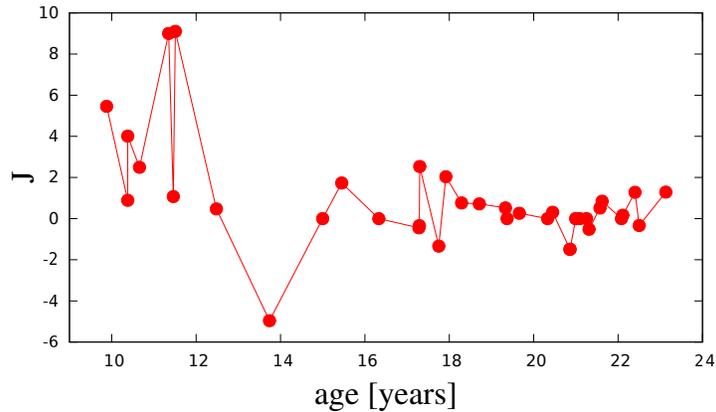}
\caption{The coefficient $J$, calculated in Eq. \ref{eq9}, against the mean age of children in the class. This coefficient is a measure of how the surplus of males in one 'camp' is correlated with a surplus of females in the other 'camp' in the same class.}
\label{fig:j}
\end{center}
\end{figure}

In Fig.\ref{fig:j} we show the contributions $J(p)$ to this average, as dependent on the age of children in the class. The semi-quantitative (and not just quantitative) character of this evaluation is due to the fact that the standard deviations $\sigma (p)$ are very different from one class to another. Were $\sigma (p)$ the same, $J(p)$ could be interpreted as just the contribution from the class $p$ to the correlation function between the number of girls in one group and the number of boys in another group in the same class. Yet, $J(p)$ is the same for both parts of the class, and it depends on $k$ and $m$ in the same way. In Fig.\ref{fig:j} we show the contributions $J(p)$ to this average, as dependent on the age of children in the class. Clearly, $J(p)$ is particularly large in some younger classes.\\

We can also ask how the fact of being female (male) influences being the member of this or that group. To check this, we have applied the contingency test \cite{cmh,brandt}: for each class $p$, we construct the $2\times 2$ contingency table Tab. \ref{tab}.

\begin{table}[!htpb]
\begin{center}
\begin{tabular}{|c|c|}
\hline
$m_1$ & $k_1$\\
\hline
$m_2$ & $k_2$\\
\hline
\end{tabular}
\end{center}
\caption{The $2\times 2$ contingency table for a given class.}
\label{tab}
\end{table}

Next, we calculate the values of $X^2$ as

\begin{equation}
X^2 =\frac{(k+m)(k_1m_2-k_2m_1)^2}{mk(k_1+m_1)(k_2+m_2)}
\label{eq8}
\end{equation}

and compare it with the related tabular values for one degree of freedom (Table I.7 in \cite{brandt}). In this way we test the null hypothesis that the group composition is uncorrelated with gender. Out of seven classes of children below 12 years old, we have to reject the null hypothesis with the probability $0.99$ for four classes, where $J>4$; for one class, the value of $X^2=6.56$ is almost at the limit $6.63$ for the same probability $0.99$. For the two remaining classes and all 30 classes of older children, the condition $X^2>6.63$ is not fulfilled.

To conclude this chapter, the numerical solution of the model equations (\ref{eq4}) allows to obtain an equilibrium state in the Heiderian balance for each class $p$. An index $J$ is proposed to evaluate the level of the gender segregation. The results given by this index are consistent with the results on the statistical significance of the segregation.

\section{Discussion}

Recall that the correlation $J(p)$ is a measure of how the surplus of males in one 'camp' of a class is correlated with a surplus of females in the other 'camp' in the same class. The results shown in Fig. \ref{fig:j} demonstrate, that most significant contributions to this correlation comes from the classes of the age below 12 years.  In particular, the highest two values of $J(p)$ come from two classes: in one of them, the obtained partition is just 'all 9 girls' against 'all 9 boys'. In the other, one part contains 15 girls and 1 boy, the other part 2 girls and 11 boys. Clearly, the gender criterion is crucial in both these classes. On the other hand, in the class of fifteen-year-olds, the partition was '2 girls and 5 boys' against '2 girls and 5 boys', and the related $J(p)$ is equal to zero. The outlier for the class of almost-fourteen-year-olds is the result of the partition: '10 girls and 17 boys' against 'no girls and 9 boys'.  Although the obtained $J(p)$ happens to be strongly negative, it is easy to imagine 
that in the 
group of 9 males, girls are not accepted at all; then, the criterion of gender could be vivid there as well.\\

To summarize, we propose a method to evaluate the gender separation in a given group. The input to the evaluation is the matrix of feelings of the group members about
other group members. These data can be non-symmetric and weighted. The partition into two parts is obtained by means of the set of differential equations, namely Eq. \ref{eq4}. The obtained stable partition only rarely and weakly depends on the parameter $\alpha$; the choice $\alpha$= 1/2 is reasonable. Further, a new index $J$ of the gender segregation is proposed. The results obtained with the poll data presented in Section III show that in classes of age below 12 years, the values of J are often about four or larger. This value can serve as a threshold for the semi-qualitative criterion of the gender segregation, until it is confirmed or denied by a better statistics.\\

Concluding, the model dynamics of removal the cognitive dissonance is generalized here to include the direct reciprocity. The same generalization allows to apply the model to non-symmetric relations. The formalism is applied to transform the data on sympathy and/or aversion, collected in 41 groups of students and teachers, into two sets in each group, with more friendly relations within the sets than between them. For each group, we evaluated the coefficient $J$, given in Eq. \ref{eq9}, as a proposed measure of gender segregation. In our data, the results indicate that the segregation is visibly stronger for the youth younger than 12 years. This conclusion is consistent with the result of psychological research, that in adolescence the preferences for same-sex friends decline \cite{pou,meht}, and it strengthens this statement by providing a semi-quantitative criterion. We hope that this work may stimulate further research on segregation in bidirectional networks in different social systems.

\appendix
\section{On the jammed states}
\begin{figure}[!hptb]
\begin{center}
\includegraphics[width=.5\columnwidth, angle=0]{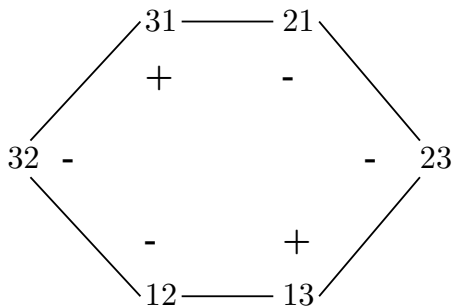}
\caption{The graph of mutual influence of the matrix elements $x_{ij}$, encoded in Eq. \ref{eq2}, for $N$=3. For example, the connection of $12$ with $13$ and $32$ means, that $dx_{12}/dt=x_{13}x_{32}$. Therefore, in stable configurations $sign(x_{ij})=sign(x_{ik}x_{kj})$. Hence either the sequence of signs is (+,-,-,+,-,-), as shown in the graph, or all signs are positive.}
\label{fig:h1}
\end{center}
\end{figure}

It is easy to prove, that for $N=3$ there is no jammed states; all stable states are balanced. The proof is graphical: only for the purpose of this proof we construct a network of six nodes, each node represents a link in the social network of $N=3$ agents, and each link between a pair of nodes means that the time evolution of each node depends on the other node. For $N=3$, this relation is reflexive. As shown in Fig. \ref{fig:h1}, the network is a cycle. Further, in the stable state the sign of each node is a product of the signs of its two  neighbors; then positive nodes have neighbors of the same sign, while negative neighbors have neighbors of different signs. Once two positive nodes are neighbors, all nodes must be positive. The only alternative is the sequence (+,-,-,+,-,-),  which can start anywhere. Yet, as seen in Fig. \ref{fig:h1}, a path between $x(i,j)$ and $x(j,i)$ contains two nodes for each pair $i,j$. On the other hand, three steps along the above given sequence leaves us at a node of the 
same sign as at the start.  Concluding, the matrix $x(i,j)$ is symmetric, {\it q.e.d.}\\

\begin{figure}[!hptb]
\begin{center}
\subfloat[One agent is liked by all other agents, yet he/she dislikes them; the relations are unreciprocated. All remaining relations are friendly.]{
\begin{minipage}{.4\columnwidth}
\vspace{-2.3cm}
\setlength{\tabcolsep}{.2cm}
\begin{tabular}{|*5{c|}}
\hline
0&-1&-1&-1&-1\\\hline
1&0&1&1&1\\\hline
1&1&0&1&1\\\hline
1&1&1&0&1\\\hline
1&1&1&1&0\\\hline
\end{tabular}
\end{minipage}
\hspace{.71cm}\includegraphics[width=.25\columnwidth, angle=0]{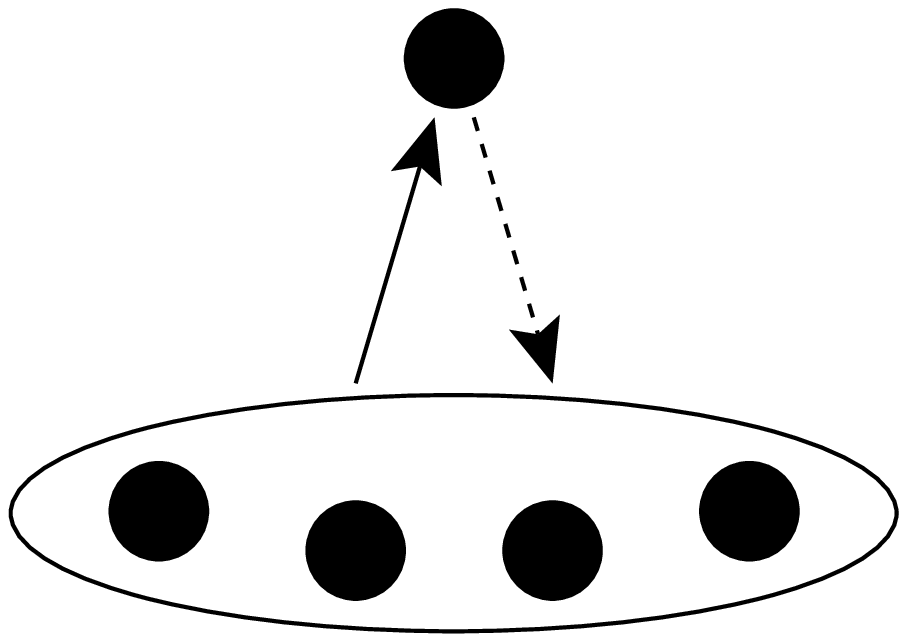}\label{fig:h2a}
}

\subfloat[One agent is disliked by all other agents, yet he/she likes them; the relations are unreciprocated. All remaining relations are friendly.]{
\begin{minipage}{.4\columnwidth}
\vspace{-2.3cm}
\setlength{\tabcolsep}{.23cm}
\begin{tabular}{|*5{c|}}
\hline
0&1&1&1&1\\\hline
-1&0&1&1&1\\\hline
-1&1&0&1&1\\\hline
-1&1&1&0&1\\\hline
-1&1&1&1&0\\\hline
\end{tabular}
\end{minipage}
\hspace{.71cm}\includegraphics[width=.25\columnwidth, angle=0]{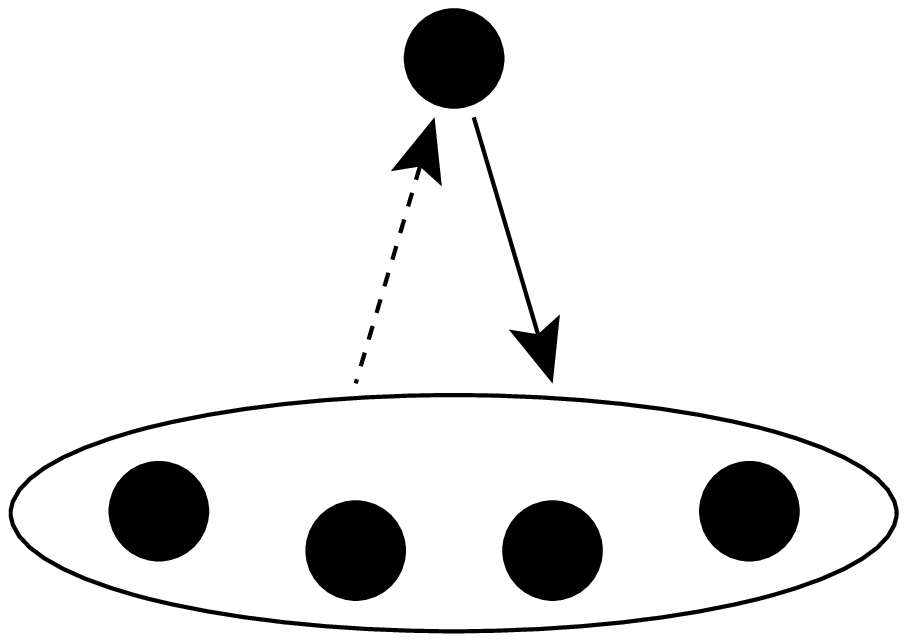}\label{fig:h2b}
}

\subfloat[Two agents remain out of the internally friendly group of three agents. The relations of one of the two (on the left) with the group of three are hostile, but his/her relations with the other one (on the right) are unreciprocated, and so are the relations of the latter with the group of three.]{
\begin{minipage}{.4\columnwidth}
\vspace{-2.3cm}
\hspace{-.53cm}
\setlength{\tabcolsep}{.18cm}
\begin{tabular}{|*5{c|}}
\hline
0&1&-1&-1&-1\\\hline
-1&0&1&1&1\\\hline
-1&-1&0&1&1\\\hline
-1&-1&1&0&1\\\hline
-1&-1&1&1&0\\\hline
\end{tabular}
\end{minipage}
\hspace{.71cm}\includegraphics[width=.2\columnwidth, angle=0]{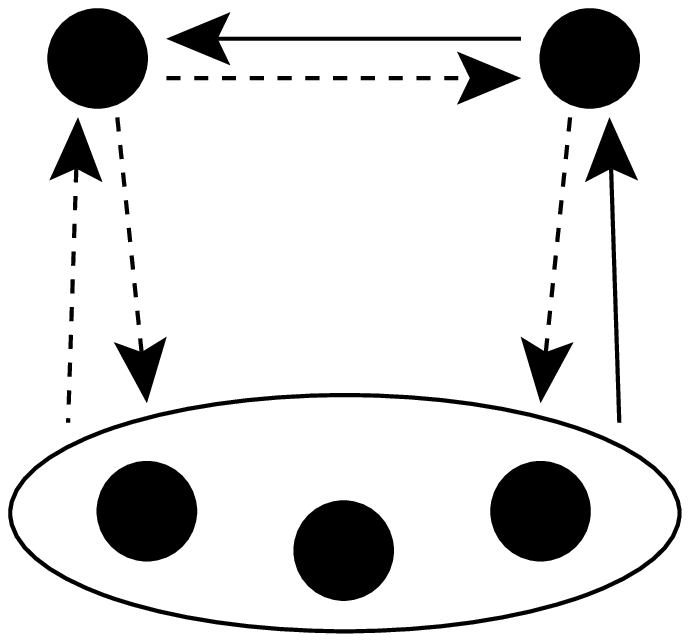}\label{fig:h2c}
}

\subfloat[Two pairs of friends are mutually hostile. The balance is destroyed by the unreciprocated relations of the fifth agent with both pairs.]{
\begin{minipage}{.4\columnwidth}
\vspace{-2.3cm}
% \hspace{-.51cm}
\setlength{\tabcolsep}{.18cm}
\begin{tabular}{|*5{c|}}
\hline
0&1&-1&-1&-1\\\hline
1&0&-1&-1&-1\\\hline
1&1&0&-1&-1\\\hline
-1&-1&1&0&1\\\hline
-1&-1&1&1&0\\\hline
\end{tabular}
\end{minipage}
\hspace{.71cm}\includegraphics[width=.25\columnwidth, angle=0]{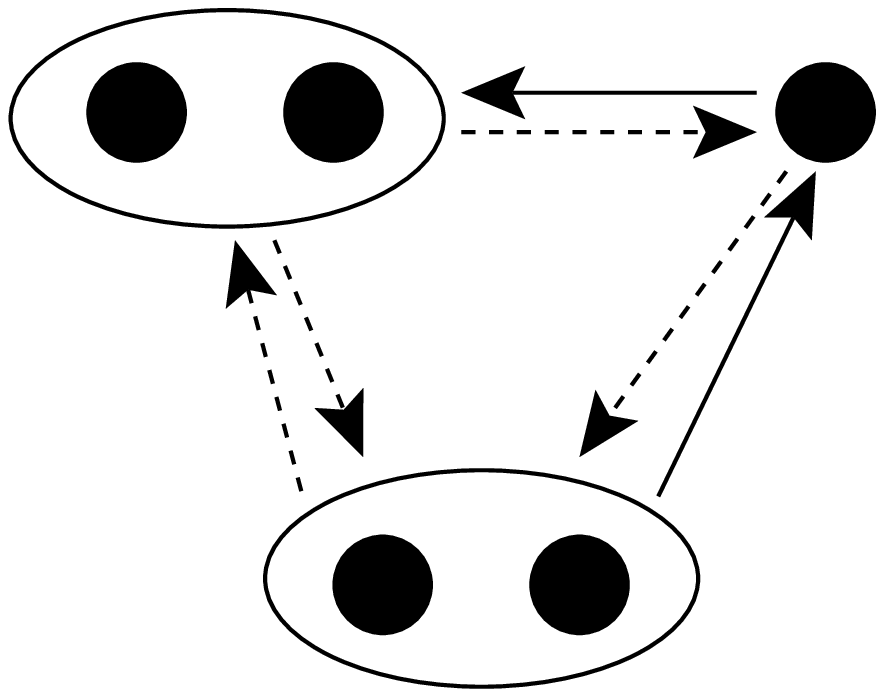}\label{fig:h2d}
}
\caption{Four patterns of stable relations where the balance is broken, for $N$=5. Dots in an ellipse mean a group of friends, with the same relations with the outsiders marked by separate dots. A friendly relation of X to Y is a continuous arrow from X to Y; a dotted arrow means a hostile relation. For each pattern, the matrix of relations is shown. }
\label{fig:h2}
\end{center}
\end{figure}

For $N=5$, the related network consists of 20 nodes, and its construction is less simple. We proved by a numerical inspection, that only 96 states out of $2^{20}$ are stable, yet only 16 out of those 96 are balanced. The remaining 80 states are jammed, i.e. stable and unbalanced; their typology reveals that there are only four types of such states, as shown in Fig. \ref{fig:h2}. Within each type, the states differ just in the order of labels of the nodes. Concluding, for the non-symmetric matrices $x(i,j)$ jammed states do exist. If an initial state is chosen sufficiently close to such a state, the system is supposed to rest there.\\

\vspace{1cm}
%% ===========================================================================
\noindent
{\bf Acknowledgement}\\
We thank Alel{\'\i} Villaverde, Sharon  Valverde, Jorge Alberto Olgu{\'\i}n and Julio C\'esar Rub\'en Romo for their technical help in gathering and processing the surveys, Wojciech S{\l}omczy\'nski and Piotr Ko\'scielniak for helpful comments. We acknowledge the partial financial support provided by DGAPA-UNAM, M\'exico, through Grant No. IN105814. The work was partially supported also by the Polish Ministry of Science and Higher Education and its grants for Scientific Research and by the PL-Grid Infrastructure.
%% ===========================================================================

%%%%%%%%%%%%%%%%%%%%%%%%%%%%%%%%%%%%%%%%%%%%%%%%%%%%%%%%%%%%%%%%%%%%%%%%%%%%%%

\end{document}